\documentclass[%
 reprint,
 amsmath,amssymb,
 aps,
]{revtex4-2}
\usepackage{graphicx}% Include figure files
\usepackage{dcolumn}% Align table columns on decimal point
\usepackage{bm}
\usepackage{color} % For comments
\usepackage{cancel}
\usepackage{array}
\usepackage{tabularx}
\usepackage{dcolumn}% Align table columns on decimal point
\usepackage{braket}
\makeatletter
\newcommand{\tpmod}[1]{{\@displayfalse\pmod{#1}}}
\makeatother

\begin{document}

\title{Topological bands in the PdSe$_2$ pentagonal monolayer }

\author{Sergio Bravo}
\email{sergio.bravoc@.usm.cl}
%\affiliation{
\address{Departamento de F\' isica, Universidad T\'ecnica Federico Santa Mar\' ia, Valpara\' iso, Chile}

\author{M. Pacheco}
%\email{monica.pacheco@usm.cl}
%\affiliation{
\address{Departamento de F\' isica, Universidad T\'ecnica Federico Santa Mar\' ia, Valpara\' iso, Chile}

\author{J.D. Correa}%
%\affiliation{
\address{Facultad de Ciencias  B\'asicas, Universidad de Medell\' \i n, 
Medell\' \i n, Colombia}

\author{Leonor Chico}
%\affiliation{
\email{leochico@ucm.es}
\address{Departamento de Física de Materiales, Facultad de Ciencias Físicas, Universidad Complutense de Madrid, 
28040 Madrid, Spain}

%\affiliation{%

\date{\today}

\begin{abstract}

The electronic structure of monolayer pentagonal palladium diselenide (PdSe$_2$) 
is analyzed from the
topological band theory perspective. Employing first-principles calculations, effective models 
and symmetry indicators, we find that the low-lying conduction bands are
topologically nontrivial, protected by time reversal and crystalline symmetries. Numerical 
evidence supporting the nontrivial character of the bands is presented. Furthermore, we obtain a relevant physical 
response from the topological viewpoint, such as 
the spin Hall conductivity.

\end{abstract}


\maketitle

\section{\label{sec:level1} Introduction \protect\\}

Two-dimensional (2D) materials are among the most promising types of systems in the
continuous search for novel structures that can give shape to future technological advances. 
The low-dimensional character of these structures makes them ideal candidates for their application 
in nanoscale devices \cite{Fiori_2014}. 
Starting with graphene more than a decade ago, a vast amount of these layered systems 
has been proposed \cite{Pere2014,Das_2015_ARCM,Novoselov_2016}. 
Within this emerging group of novel 2D systems, pentagonal 2D materials are attracting increasing attention because of their symmetry. For instance, the proposal of pentagraphene stirred much attention \cite{2014_tang,2015_Zhang_pentagraphene,Bravo_SR_2018,Correa_2020}. Furthermore, the possibility of presenting topologically nontrivial phases has driven the interest to other pentagonal layers with different compositions \cite{2016_TiC2,Zhuang2019,Bravo_SR_2019,Bravo_NS_2021}. However, some theoretical proposals have been shown to be structurally unstable, specially those with  dominant p-orbital bonds \cite{Avramov2015,Kuklin2020}. Notwithstanding, there are several instances of experimentally found 2D pentamaterials, such as PdSe$_2$, PdS$_2$ and NiN$_2$ \cite{oyedele_pdse2_2017,Zhang2021,Bykov2021}.
 Among these, the recently
synthesized PdSe$_2$ has been the subject of intensive experimental and theoretical research. 
Importantly, the pentagonal phase is the stable allotrope for this material. Different forms of pentagonal PdSe$_2$ with a variable number of layers have been reported, possessing high
air stability \cite{oyedele_pdse_2017,nguyen_atomically_2020,PdSe2_defects,PdSe2-air-stab,PdSe2_ACSnano}, 
which is an essential characteristic for their extended (long-term) use. 
Also, electrical transport characteristics \cite{oyedele_defect-mediated_2019}, 
remarkable optical \cite{PdSe2_lindic,PdSe2_nonlinopt,PdSe2_optoel,PdSe2_nonlin_Multilayer},
and good thermoelectric properties \cite{PdSe2_thermoelec} have been experimentally reported. 
On the theoretical side, several works have analyzed the physical properties of the 
material in its monolayer and multilayer form
\cite{PdSe2_dft2015,PdSe2_renorm-gap,PdSe2_defectsPRB,PdSe2_compress,PdSe2_NJP_bandgap}. 
Among these theoretical accounts the band connectivity and the symmetry-related properties 
of the electronic structure have not been studied in detail for monolayer PdSe$_2$; we address this
issue in this work. Using the theory of symmetry indicators along with first-principles 
calculations, we identify that the lowest conduction bands of monolayer PdSe$_2$ realize a topologically
nontrivial phase. These bands comprise a strong topological phase with a well defined topological 
invariant and gapless edges states that we characterize using well-known numerical methods. Also, an analysis 
on the accessibility of these nontrivial conduction states by Fermi level manipulation
is presented. These results open the possibility for the exploration of this promising material 
and its related structures from the topological point of view.

The article is organized as follows. First, an overview of the numerical calculations and parameters 
used is sketched. This is followed by 
the geometric information and first-principles electronic band 
structure of monolayer PdSe$_2$, along with a study of the symmetry character of the bands. 
Additionally, effective models based on Wannier interpolation are briefly described and put forward to 
study the edge states and the Wannier charge center (WCC) evolution along different directions. 
The spin Hall conductivity, a signature of its nontrivial band character,
is computed as a function of frequency and chemical potential. Finally, we 
conclude with a summary and outlook for possible future avenues to explore the potential of this material. Part of our results are left 
as Electronic Supplementary Information (ESI)$^\dagger$.  

\section{Computational details}\label{II}
 The calculations for the band structure were carried
with standard density functional theory (DFT) method using the QUANTUM ESPRESSO (QE) package \cite{QE_2020}
at the GGA (generalized gradient approximation) level within the Perdew-Burke-Ernzerhof (PBE) implementation. We have also corroborated the robustness of our main results by resorting to the GPAW code \cite{mortensen_real-space_2005,enkovaara_electronic_2010} and checking for several DFT functionals; these additional computations are summarized in Appendix A. 
The monolayer structure was relaxed with a force tolerance of $10^{-4}$ eV/\AA.
The energy cutoff for the plane wave basis was 100 Ry with a vacuum distance of 20 \AA \hspace{0.01cm} 
in the perpendicular direction to the monolayer. A Monkhorst-Pack grid of $15\times 15 \times1 $ was 
chosen and the energy convergence tolerance was set to $10^{-8}$ Ry. 
The Wannier interpolation of the DFT energy bands was performed with the Wannier90 
code \cite{mostofi_updated_2014}. 
Two models were implemented; a twelve-band (12B) model including the four uppermost valence band 
and the eight lowest conduction bands, and an  eight-band (8B) model that only includes the above-mentioned conduction bands. This latter model was used to focus only on the nontrivial bands 
of the system. 
For the 12B model, $d$-orbitals were used for the Pd atoms and $p$-orbitals for the Se atoms.
These orbitals were only used as starting sites for the orbitals, since the location of the Wannier centers 
may change under the wannierization process. The post-processing of Wannier-based models was
carried by the PythTB code \cite{Pythtb}. Also, Wannier90 was used to analyze this model and to calculate the optical responses presented below. The mathematical expressions implemented in 
this code are based on Refs.~\cite{ibanez-azpiroz_ab_2018,SHC_w90} and for quick reference 
are presented in the ESI.$^\dagger$ 

\section{Electronic band structure and symmetry indicators}\label{III}

\subsection{Lattice geometry and space group}\label{R1}
\begin{figure}
\centering
\includegraphics[width=1.\columnwidth,clip]{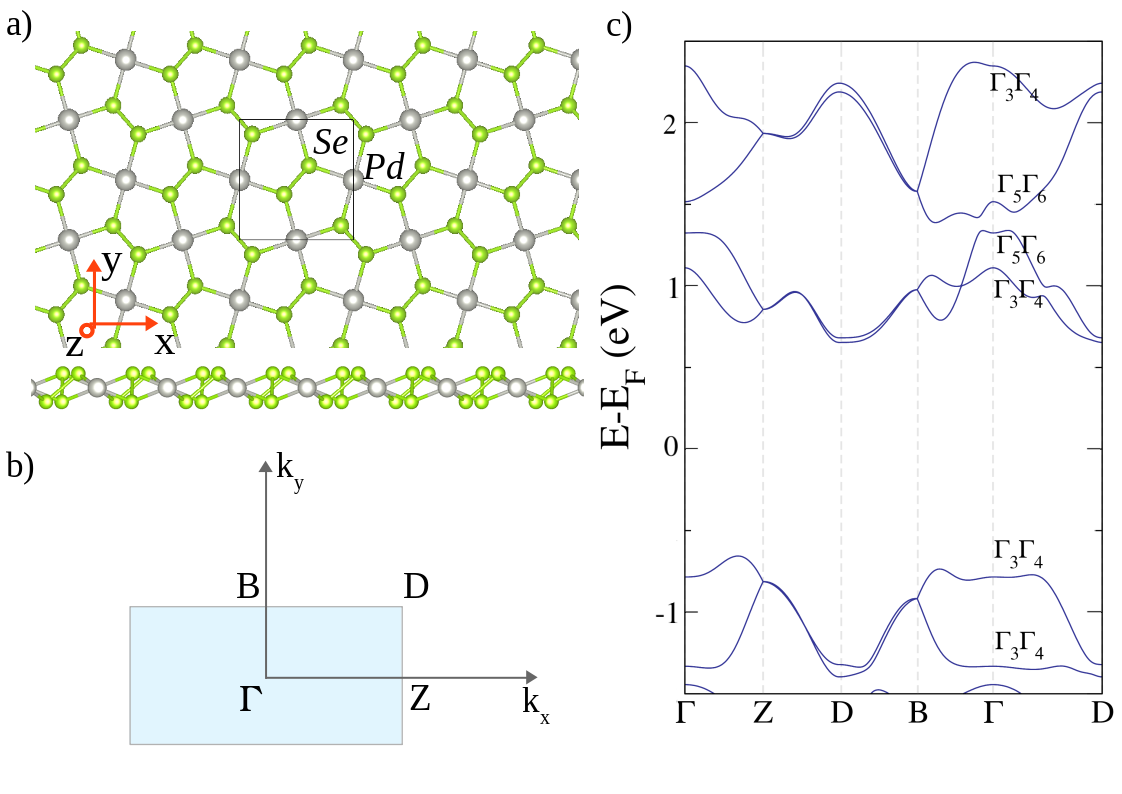}
\caption{a) Lattice structure of monolayer PdSe$_2$. b) Brillouin zone for the space group \#14.
c) Electronic band structure of monolayer PdSe$_2$ along a high-symmetry path. Band energies are referred to the Fermi level $E_F$.}
\label{fig1} 
\end{figure}

The lattice structure of PdSe$_2$ is composed of 
irregular (type 2) pentagons forming a buckled geometry as presented in Fig.~\ref{fig1} a) \cite{ZHUANG2019448}. 
Pd atoms are fourfold coordinated and Se atoms have coordination three. As previous 
works reported \cite{oyedele_pdse_2017}, the crystalline order conforms a tetragonal 
lattice, with three symmetry operations; a twofold rotation around one of the lattice 
vectors axis with fractional translation (1/2,1/2,0) in terms of the unit cell vectors, 
a mirror glide plane and spatial inversion \cite{bradley_group}. 
This has to be complemented with time reversal (TR) symmetry to give the space
group (SG) P2$_1$c or SG \#14. 
The Pd atoms sit at the $2a$ Wickoff 
position (WP) while the Se atoms locate at the $4e$ WP \cite{bradley_group}. 
The relaxed structure obtained from first-principles calculations comprises a rectangular unit cell 
with lattice vectors with magnitude $a=5.74$ \AA \hspace{0.01cm} and $b = 5.91$ \AA. 
The calculation also yields a puckering distance 
of 0.7 \AA, confirming the buckled 
geometry of the material. These results show a good agreement with the reported experimental 
and theoretical values \cite{PdSe2_dft2015,PdSe2_NJP_bandgap,oyedele_pdse_2017,Lei-PdSe2-nanosc}.    
The band structure of monolayer PdSe$_2$ 
has been extensively studied in previous works 
\cite{oyedele_pdse_2017,PdSe2_dft2015,PdSe2_renorm-gap,PdSe2_defectsPRB,PdSe2_compress,PdSe2_NJP_bandgap,Lei-PdSe2-nanosc}. As indicated above, we have also carried out an extensive investigation of its electronic properties by employing several exchange-correlation functionals, detailed in Appendix A. We confirm the robustness of the nontrivial topology found for this material. 

Here we focus on the band connectivity an associated topological properties of the low-energy 
bands around the Fermi level. 
For this purpose we present in Fig.~\ref{fig1} c) the electronic band structure including 
spin-orbit coupling (SOC) at the PBE level along the high-symmetry path $\Gamma-{\rm Z}-{\rm D}-{\rm B}-\Gamma-{\rm D}$
in momentum space. The Brillouin zone is depicted in Fig. \ref{fig1} b), following the 
notation of Ref.\cite{bilbao_serverII}.

The inclusion of SOC  is crucial for the results obtained as will be clarified in what follows. It is 
well-known that PBE functionals systematically underestimate the fundamental gap in semiconductors and insulators. 
As stated above, we verified that other functionals yield similar results and predict the nontrivial topology found with QE, as detailed in the Appendix. 
Additionally, we have 
revised the existing literature concerning the electronic 
structure of monolayer PdSe$_2$ obtained with the 
hybrid (HSE06) functional \cite{PdSe2_renorm-gap} and at the GW level \cite{GW_PdSe2}. 
These more expensive calculations show a larger band gap but also keep the band general features 
unaltered, giving validity for the topological analysis using the PBE functional presented here. 

From the band structure and the space group information it can be confirmed that all bands are doubly
degenerate along the whole Brillouin zone (BZ) since  monolayer PdSe$_2$ is centrosymmetric (SG\#14). 
The spin-orbit interaction - which requires the use of double space groups - obviously affects the degeneracies of the system without SOC, only leaving the possibility of fourfold nodal points at the Z and B points. 
These points are protected by the nonsymmorphic symmetries of the SG in conjunction with time-reversal
symmetry \cite{dresselhaus_group_2008}, and they are present in every group of
four bands in the structure. This is the basic ingredient for the band connectivity of the system, 
since 
these sets of four bands form a band representation \cite{TQC_PRB}, following the prescription of topological quantum chemistry theory and symmetry-based indicators \cite{TQC_rev,Kruthoff,Po2017}. 

The first step in the 
study of the topological properties 
of monolayer PdSe$_2$ 
is the identification 
of nontrivial topology signatures in the corresponding space group, namely, SG \#14. 
Firstly, it is customary to look for strong topology,
since this is the most widely known nontrivial phase. 
To this end, we employ the results of 
Ref.~\cite{Elcoro2020_smith}, adapted for the case of a two dimensional BZ.  
Specifically, a Smith normal form decomposition is applied to the set of elementary 
band representations (EBR) of the SG. Recall that EBR are the building blocks to construct the bands of 
atomic insulators and as such, they can be mapped directly to exponentially localized Wannier 
functions \cite{vanderbilt_2018} situated at the atomic positions of the material. 
The procedure is briefly sketched as follows. First an EBR matrix is constructed, including as 
coefficients the multiplicities of the irreducible representations (IR) at the high-symmetry 
points of the material ($\Gamma$, Z, D and B for our particular case). 
The elements of the EBR matrix are provided in Table \ref{tab1}
(using the Bilbao Crystallographic Server information \cite{bilbao_serverII}).
Next, the Smith normal form matrix $\Delta$ is calculated, which is a diagonal matrix with positive 
integer values. If some of these values are greater that one, then a strong topological phase is 
possible for the SG \cite{Elcoro2020_smith}. The $\Delta$ matrix for the (2D) SG\#14 is given by 
\begin{equation}
\Delta =\begin{bmatrix}
1 & 0 & 0 & 0 \\
0 & 1 & 0 & 0 \\
0 & 0 & 2 & 0 \\
0 & 0 & 0 & 0 \\
0 & 0 & 0 & 0 \\
0 & 0 & 0 & 0 \\
0 & 0 & 0 & 0 \\
0 & 0 & 0 & 0 
\end{bmatrix}.
\end{equation}

\begin{table}[ht]
\begin{tabular}{l|cccc}
IR/EBR       & \multicolumn{1}{l}{EBR1} & \multicolumn{1}{l}{EBR2} & \multicolumn{1}{l}{EBR3} & \multicolumn{1}{l}{EBR4} \\ \hline
$\Gamma_{3}\Gamma_{4}$ & 2                        & 2                        & 0                        & 0                        \\
$\Gamma_{5}\Gamma_{6}$ & 0                        & 0                        & 2                        & 2                        \\
D$_3$          & 1                        & 0                        & 1                        & 0                        \\
D$_4$          & 1                        & 0                        & 1                        & 0                        \\
D$_5$          & 0                        & 1                        & 0                        & 1                        \\
D$_6$          & 0                        & 1                        & 0                        & 1                        \\
Z$_2$          & 1                        & 1                        & 1                        & 1                        \\
B$_2$          & 1                        & 1                        & 1                        & 1                       
\end{tabular}
\caption{Elements of the EBR matrix for the two-dimensional SG\#14.}
\label{tab1}
\end{table}

\begin{figure}
\centering
\includegraphics[width=0.95\columnwidth,clip]{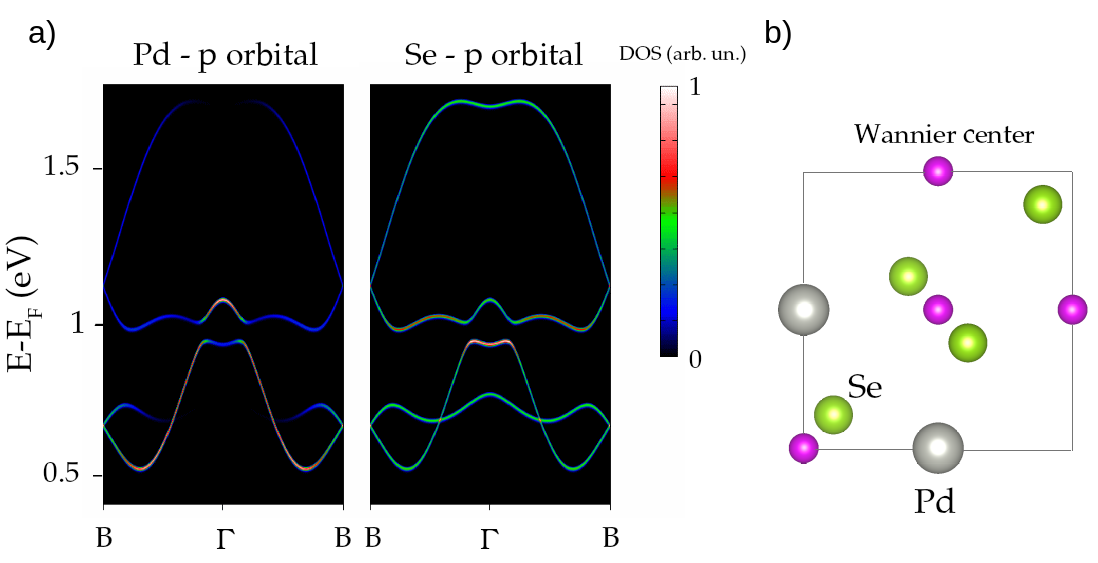}
\caption{a) Orbital-projected bands for monolayer PdSe$_2$. b) Relative position of the Wannier centers within the unit cell of monolayer PdSe$_2$.}
\label{fig2} 
\end{figure}

We observe that a diagonal element with value 2 is present, which allows for a $\mathbb{Z}_2$ phase in the 
system \cite{Elcoro2020_smith}. Thus, strong topological bands are in principle possible in this SG. 
This phase is a version of the time-reversal plus inversion topological phase \cite{Alex2014}, which 
in this case is further enriched by the other crystalline symmetries of the group. The presence of 
these additional symmetries simplifies the symmetry-indicated 
character of the nontrivial topology. 
This can be seen in the definition of the above mentioned $\mathbb{Z}_2$ invariant. 
Further manipulation of the EBR matrix (see the ESI$^\dagger$ for this particular group or Ref.~\cite{Elcoro2020_smith} 
for the general theory) gives as a result that the $\mathbb{Z}_2$ invariant only depends on the parity of 
the $\Gamma$ point IR. In this double SG $\Gamma$ only has two IR, $\Gamma_{3}\Gamma_{4}$ and
$\Gamma_{5}\Gamma_{6}$ \cite{bilbao_doubleSG}. With this definition the topological invariant 
for the strong phase can be defined as   
\begin{equation}
      \mathbb{Z}_2 = n_{\Gamma_{3}\Gamma_{4}} \; ({\rm mod} \; 2)  = n_{\Gamma_{5}\Gamma_{6}}\; ({\rm mod} \; 2) ,
\end{equation}
 where $n_{\Gamma_{3}\Gamma_{4}},n_{\Gamma_{5}\Gamma_{6}}$ correspond to the multiplicities of the IR at $\Gamma$. 
In other words, if a single (fourfold) band representation or a group of band representations has an odd 
number of $\Gamma_{3}\Gamma_{4}/\Gamma_{5}\Gamma_{6}$ IR, then these bands are topological. This definition
of $\mathbb{Z}_2$ is simpler than the standard Fu-Kane formula for TR inversion-symmetric insulators \cite{Fu-kane2007} 
due to the additional constraints of the above-mentioned symmetries.
Note that 
this is the only kind of symmetry-indicated topology that can be present in this two-dimensional (layer) version of SG\#14. 
The linear combination of strong band representations may yield either a strong band or 
a trivial band representation. This differs from the three-dimensional version of this group, where fragile
bands and strong bands can coexist \cite{Elcoro2020_smith}.  
We have numerically computed the IR characters for monolayer PdSe$_2$ from the first-principles 
electronic structure using the IrRep package \cite{iraola2020irrep}. We consider two groups of 
bands: the valence band manifold of the material and the eight lowest conduction bands. In the
valence band set there exists an even number of $\Gamma_{3}\Gamma_{4}/\Gamma_{5}\Gamma_{6}$ irreducible representations,
which render the material a trivial insulator.
On the other hand, if we take into account the aforementioned conduction bands, we find that both fourfold band 
representations have separately an odd number of $\Gamma_{3}\Gamma_{4}/\Gamma_{5}\Gamma_{6}$ IR. 
Therefore, each single group realizes a strong topological group of bands. In Fig.~\ref{fig1} c) we have labeled %the bands for 
the highest valence band 
and the lowest 
conduction bands  
with the corresponding IR. 
A crucial role here is played by the SOC, which permits a band inversion between the two group 
of band representations. The inversion can be initially identified from the band structure 
in Fig.~\ref{fig1} c), specially along the ${\rm B}-\Gamma$ line. 
Further confirmation of this band inversion is presented in Fig.\ref{fig2} a), where the 
orbital-projected bands are presented. The $p$-orbitals constitute the most important contribution to
the inversion and the effective models for these bands are based on this result.        

\subsection{Evidence of the nontrivial topology}\label{R2}
\begin{figure}
\centering
\includegraphics[width=1.\columnwidth,clip]{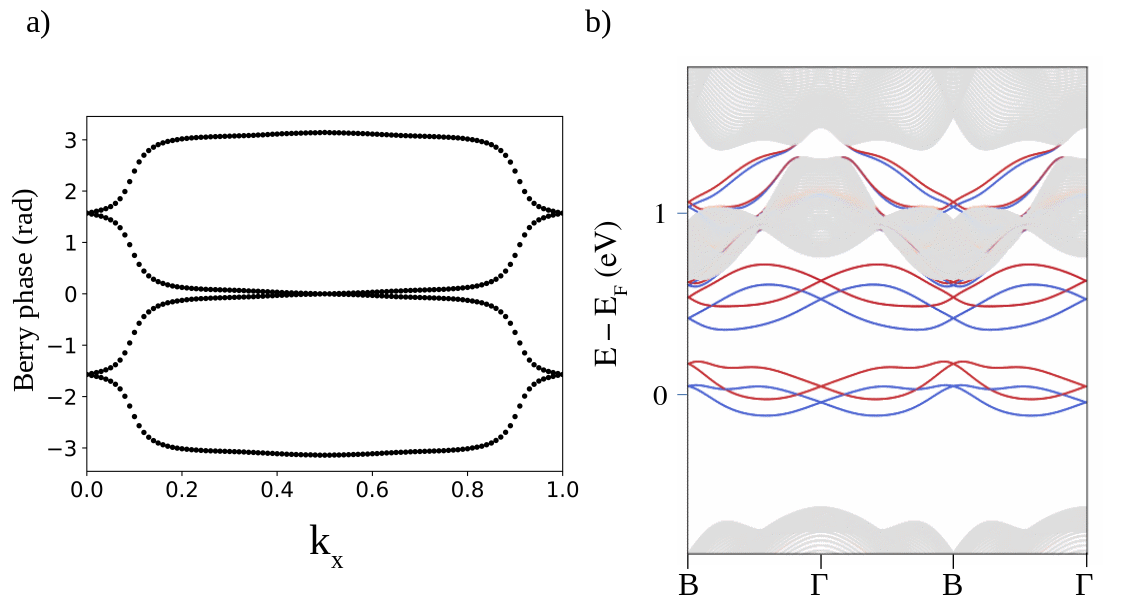}
\caption{a) Evolution of the Wannier charge center (Wilson loop) along the $k_x$ direction. 
b) Energy dispersion of the ribbons for the Wannier interpolated 12B model. Red-colored and blue-colored bands represent
states that are localized at the upper and lower edges of the ribbon, respectively.}
\label{fig3} 
\end{figure}

 A clearer picture of the nontrivial topology of monolayer PdSe$_2$ can be obtained by the computation of the standard quantities that pinpoint the nontrivial character of the bands. 
Thus, we calculated the energy dispersion for a ribbon geometry of the material and the bulk Wannier charge center
(WCC) evolution along the BZ \cite{wccPRB}, using a Wannier-based model.
Before delving into these results, we would like to comment on the Wannier centers in real space associated with the conduction bands. 
We employed a reduced 8B model, as described above, to account only for these bands. They are well 
separated from the higher conduction bands, which make them suitable for faithful wannierization. The localization of the
Wannier centers in real space for PdSe$_2$ is represented in Fig.~\ref{fig2} b). A pair of Wannier centers sits on each 
site, as dictated by time reversal symmetry (Kramers pairs) \cite{soluyanovZ2}. Most importantly, it can be 
observed that two pairs of these Wannier centers are localized at a WP (2b) that is not occupied by any of 
the atoms in the material. 
This $\textit{obstruction}$ hints for a nontrivial topology \cite{soluyanovZ2} and PdSe$_2$ can be dubbed as a
conduction-band-obstructed atomic insulator. The other Wannier centers localize on the 2a WP and thus coincide with atomic (Pd) orbitals.  
The WCC evolution along the $k_x$ direction in the BZ is presented in Fig\ref{fig3} a). Here we also use the 
8B model in order to isolate the nontrivial behavior. For this model we assume that the four lowest conduction bands
are occupied, just 
to conform with the usual WCC definition \cite{vanderbilt_2018}. 
The general trend of the WCC evolution shows the typical features of a TR inversion-symmetric 
topological insulator \cite{Alex2014}, with the nontrivial crossings at one of the time-reversal invariant 
momenta (in this case $k_x = \pi$). 
The energy dispersion of the edge states for a confined geometry is plotted in Fig.~\ref{fig3} b). In this case we
make use of the 12B model, with the aim to show the fundamental gap and their edge states. Gapless energy 
states arise within the gap at $\approx 
1.3$ 
eV 
above the Fermi level, localized on the edges of the finite slab, giving
further confirmation for the nontrivial phase of the conduction bands. The edge termination has influence on the %form of 
dispersion of the edge states, but there are  always gapless states in this upper gap. The fundamental
gap also presents edge states that are trivial in terms of the above classification. Further information for other
edges is presented in the ESI$^\dagger$.

\section{Topologically relevant physical response}\label{IV}
 The main drawback of the topological bands spotted in the 
above discussion is that they are situated in the conduction bands, which implies that some external 
manipulation of the material is necessary in order to access them. 
This hinders the use of monolayer PdSe$_2$ as a spin Hall insulator, since the nontrivial gap is not
the fundamental gap. 
Notwithstanding, the lowest conduction bands of the material are reachable by standard doping and gating \cite{Wang_2012,Fiori_2014}. 
Nowadays there are gating techniques for 2D materials which allows for higher doping levels than standard methods \cite{Bisri2017,Morpurgo2021}.
As these procedures can displace the chemical 
potential $\mu$ to lie {\it inside} the conduction bands, it is
possible to
access the nontrivial bands and explore the nontrivial behavior of the material.
We simulate this 
electron doping effect by a rigid shift of the chemical potential using the 12B model
and calculate an optical response, namely,  
\begin{figure}
\centering
\includegraphics[width=1.\columnwidth,clip]{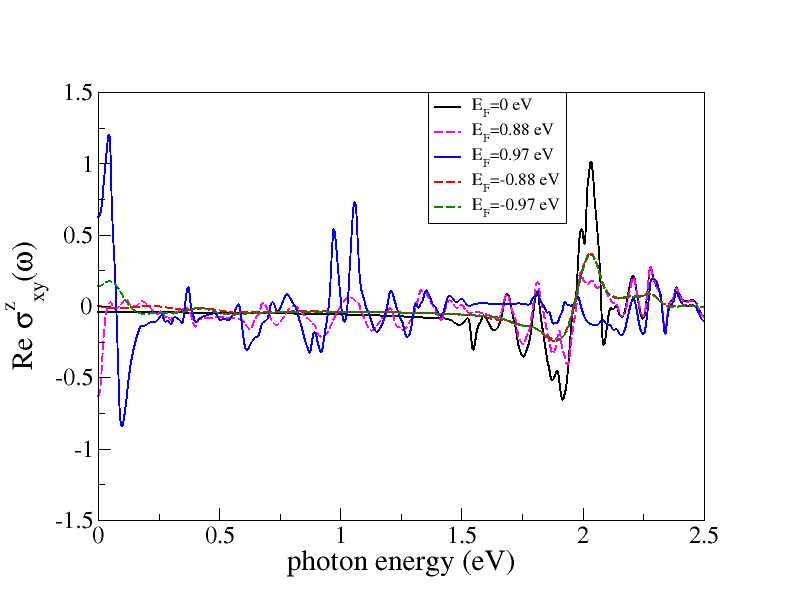}
\caption{Real part of the spin Hall conductivity $\sigma^{z}_{xy}$ as a function of external photon energy for different chemical potential values}
\label{fig4} 
\end{figure}
the frequency-dependent spin Hall conductivity (SHC) for different values
of the chemical potential. The slab geometry has been taken into account 
by a global scaling factor following Ref.~\cite{ibanez-azpiroz_ab_2018}. 
We report the real part in Fig.~\ref{fig4}; only the usual component 
with 
spin along the $z$ direction and transverse 
current with respect to the applied external field is presented. 
We have calculated cases with the chemical potential within the conduction band and also inside 
the trivial valence bands to assess the magnitude of the response. 
Additionally, the SHC with no doping effect is included. It can be appreciated that both 
types of doping produce a new intense peak in the low-frequency range of the spectrum. 
However, when $\mu$ is situated around the nodal features of the conduction bands - around 1 eV 
above the original Fermi level - the magnitude of this peak is greatly enhanced,  
 giving 
additional support to the nontrivial character of 
the bands. Although the topological character is not
essential for an enhanced magnitude of the effect, 
as can be 
checked by the large static SHC in Pt 
\cite{SHC_Pt_PRL}, it can boost the optical response. 
Several other works have reported sizable values 
for two-dimensional semiconductors 
\cite{SHC_calc123d,SHC_ingap1}.
It is important to note that the above-mentioned 
optical effect could be affected by the inclusion
of 
quasiparticle effects, such as those discussed in Ref.~\cite{PdSe2_renorm-gap} for the linear optical response.

Additionally, from Fig.~\ref{fig4} it can be 
observed that the static limit ($\omega 
\rightarrow 0$) of 
the optical SHC yields a nonzero value even for 
the undoped system, implying a nonzero SHC within
the fundamental gap. 
This situation has been encountered in previous 
calculations of the SHC on semiconductors such as 
in Refs. \cite{SHC_semic_PRL,SHC_sign}, where it 
is mentioned  that the in-gap SHC cannot give rise to spin accumulation for trivial systems.
Recently \cite{SHC_ingap1}, it has been argued 
that this SHC is only a numerical artifact due to 
the use 
of a broadening factor and, by means of degenerate perturbation theory, they obtain zero in-gap 
conductivities
for trivial insulators. For completeness we have 
computed this static SHC for monolayer PdSe$_2$, 
using the method of Ref. \cite{SHC_ingap1} as 
implemented in Ref. \cite{PAOFLOW}, and also with 
the Wannier90 
code \cite{SHC_w90}, which uses the standard 
broadening factor. 
In both cases we obtain a constant nonzero value 
along both the trivial and the nontrivial gap. 
Yet, the magnitude in the case of the trivial gap 
is low and in principle could not be 
detectable. This static SHC is presented in the 
ESI$^\dagger$. 

\section{Conclusions}\label{Conc}
 In this work
we address 
the electronic band topology of monolayer 
PdSe$_2$. We have shown that, although the valence bands of the material are
trivial and in principle the system is a trivial 
topological insulator, the lowest conduction bands have a 
nontrivial topology. By means of standard analysis we have found that the group of lowest 
fourfold-degenerated conduction 
bands are separately nontrivial, characterized by 
a $\mathbb{Z}_2=1$ invariant. This phase is strong
and gives rise to gapless
edge states in the conduction band gap and to a nontrivial WCC evolution. The nontrivial character of these bands
can be in principle accessible by doping, and we have presented numerical evidence for enhanced responses in the 
case of the
%shift current and 
optical spin Hall conductivity at low frequency.
It can be mentioned that other materials with the same pentagonal structure and space group in monolayer form 
has been theoretically reported in several computational databases \cite{haastrup_computational_2018,choudhary_high-throughput_2017}. Many of these materials - such 
as for example PtTe$_2$, PdTe$_2$ and NiTe$_2$ - show a similar behavior for the low-lying conduction 
bands and are expected to host similar nontrivial bands. Further work is also needed to elucidate 
other connections between the nontrivial bands and physical responses not covered here. Additionally, 
topological bands could likely be present in multilayer PdSe$_2$ and their effects for the low energy
regime are worth to be explored.    

\section*{Appendix: Electronic characterization 
of 
PdSe$_2$ monolayer}
\label{append}
\renewcommand{\theequation}{A$.$\arabic{equation}}
\renewcommand{\thefigure}{A$.$\arabic{figure}}  
\renewcommand{\thetable}{A$.$\arabic{table}}  
\setcounter{equation}{0}
\setcounter{table}{0}
\setcounter{figure}{0}

In this section we  present the theoretical 
electronic characterization  of  the  PdSe$_2$ 
monolayer, i.e., the electronic band structure 
computed within a first-principles approach with 
different exchange-correlation functionals.  The  
calculations are performed in  the  framework of  
density  functional  theory (DFT)  as   
implemented  in  GPAW using  several  GGA  and 
van der Waals (vdW) exchange-correlation functionals 
\cite{mortensen_real-space_2005,enkovaara_electronic_2010}. A projected  augmented wave (PAW) method was  employed for the basis set with  an  energy  
cutoff of  $650$ eV. For the $k$-points  we used  
a  grid  of $15\times 15\times 1$. All structures 
are  fully relaxed  until  the  atomic  forces in 
each atom  were less  of  $0.02 $ eV/{\AA}. 

\begin{table}
\centering
 \begin{tabular}{ccccc}
 \hline 
 XC & a(\AA) & b(\AA) & gap (eV) & group  \# \\
 \hline
 PBE & 5.84960 & 5.90720 & 1.108 & 14 \\ 
 LDA & 5.73997 & 5.78863 & 1.171 & 14 \\ 
 RPBE & 5.89355 & 5.95226 & 1.086 & 14 \\ 
 revPBE & 5.88490 & 5.93966 & 1.094 & 14 \\ 
 vdW-DF & 5.96822 & 6.04448 & 0.943 & 14 \\ 
 vdW-DF2 & 6.03519 & 6.03519 & 0.850 & 14 \\ 
 optPBE-vdW & 5.90887 & 5.98340 & 0.976 & 14 \\ 
 C09-vdW & 5.82330 & 5.89056 & 1.025 & 14 \\ 
 GLLBSC  &         &         & 1.880 &    \\
 exp-bulk\cite{oyedele_pdse2_2017} & 5.7457  & 5.8679\\
 \hline
 \end{tabular} 
 \caption{Obtained structural  parameters  and  band gap of  PdSe$_2$ monolayer  employed  different  exchange-correlation  functionals. }\label{TB0}
\end{table}

The computed values of the lattice constants and 
the bandgap of the 2D pentagonal PdSe$_2$  are 
shown in Table \ref{TB0}. Comparing to the 
experimental values for bulk PdSe$_2$ reported in 
Ref. \cite{oyedele_pdse2_2017}, we observe that 
PBE functional shows a maximum difference of 0.05 
\AA\ for the $b$ lattice parameter. If vdW 
interactions are considered, the maximum 
difference in lattice constant is 0.26 \AA\ for 
the vdW-DF2 functional, in agreement with a 
previous report of theoretical lattice parameters 
of bulk PdSe$_2$ \cite{oyedele_pdse2_2017}. 
Independently of  the geometric configuration  
obtained for every  exchange-correlation  
functional,  the  symmetry  group  of monolayer 
PdSe$_2$  is  preserved.   

\begin{figure}
\centering
\includegraphics[width=0.9\columnwidth,clip]
{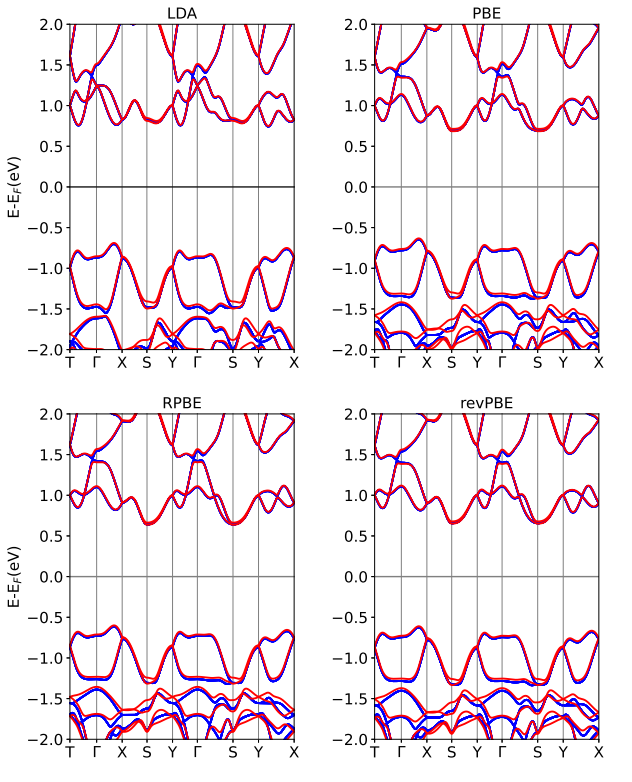}
\caption{Band structures  of  PdSe$_2$ monolayer. 
Each panel  represents  a different  GGA 
exchange-correlation  functional.  Red   bands  
include SOC; blue bands are computed without SOC.}
\label{ES_S1} 
\end{figure}

\begin{figure}
\centering
\includegraphics[width=1\columnwidth,clip]{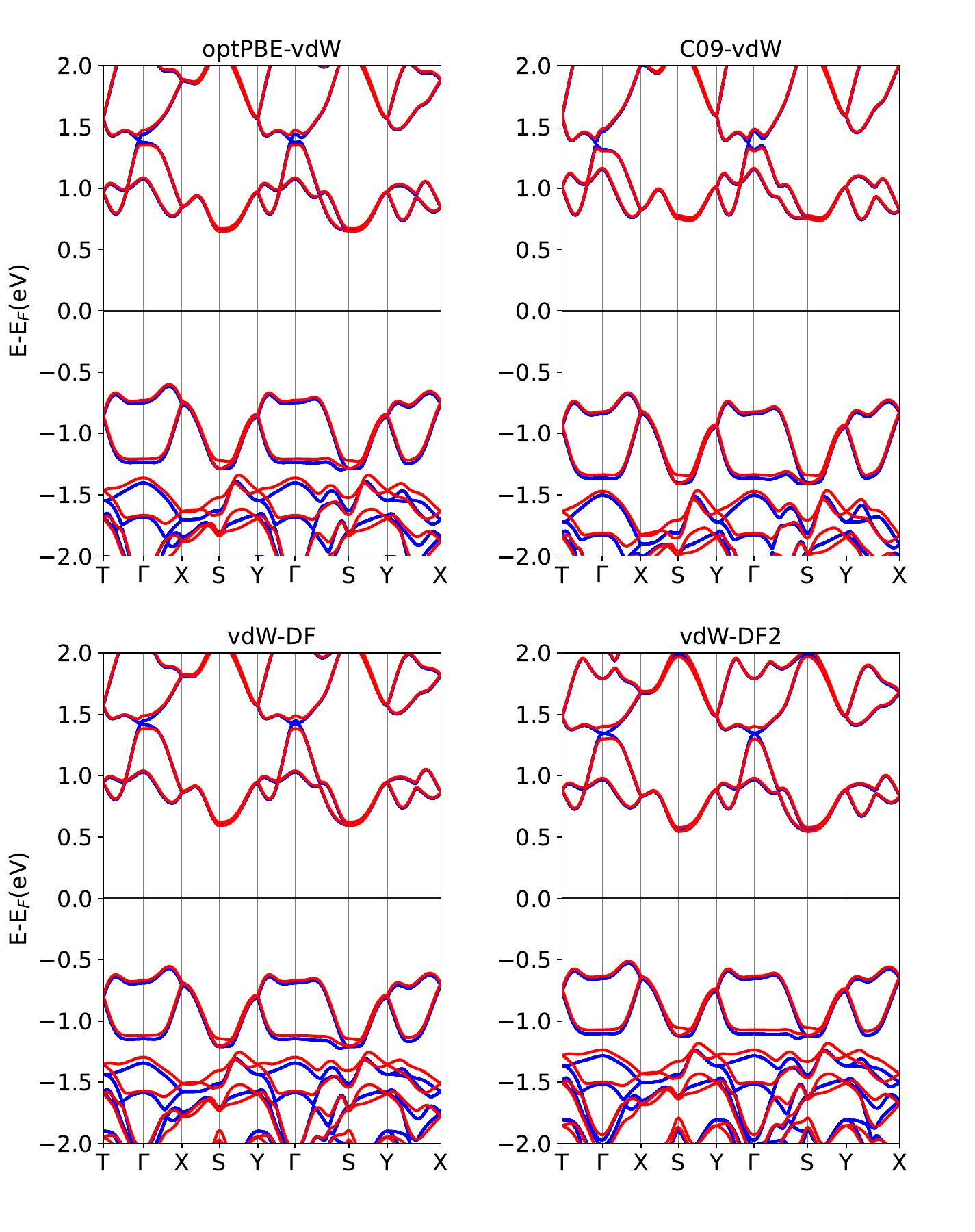}
\caption{Band structures  of  PdSe$_2$ monolayer. Each panel  represents  a different  vdW  functional.   Red   bands  include SOC; blue bands are computed without SOC.}
\label{ES_S2} 
\end{figure}
The  band structures  for  PdSe$_2$ monolayer are 
shown in  Figs. \ref{ES_S1} and  \ref{ES_S2}. The 
$k$-point path is labeled using the high-symmetry 
points of the of the 3D orthorhombic lattice. Our 
results  show  that,  independently  of  the 
exchange-correlation  functional,  the  topology 
characteristics of  the  bands  are  preserved and
are the same as those obtained with QE 
calculations.

\section*{Acknowledgements}
This work has been partially supported by Chilean FONDECYT Grant 1211913, Spanish MCIU and AEI and the European
Union under Grant No. PGC2018-097018-B-I00 (MCIU/AEI/FEDER, UE), and by Grant USM-DGIIP PI-LI 1925. 

\newpage
\bibliography{main}
%\bibliography{Ref_misc,Ref_PdSe2,Ref_optics,Ref_SHC,Ref_JDC}

\end{document}

\author{\text{Sergio Bravo}$^\dagger$}
\author{M. Pacheco}
%\author{\text{M. Pacheco}$^\mathsection$}
\affil{Departamento de F\' isica, Universidad T\'ecnica Federico Santa Mar\' ia, Valpara\' iso, Chile}

\author{J.D. Correa}%
\affil{Facultad de Ciencias  B\'asicas, Universidad de Medell\' \i n, 
Medell\' \i n, Colombia}

\author{Leonor Chico}%$^\star$}%
\affil{GISC, Departamento de Física de Materiales, 
Facultad de Ciencias Físicas, 
Universidad Complutense de Madrid, 
28040 Madrid, Spain}
\affil[ ]{ }
\affil[$\dagger$]{sergio.bravoc@usm.cl}
%\affil[$\mathsection$]{monica.pacheco@usm.cl}
%\affil[$\star$]{leochico@ucm.es}
%\author{}
%\date{}

\maketitle

%\bigskip

\subsection*{A. Topological invariant extraction from the symmetry indicators of space group \#14}

In this section we follow Ref. \cite{Elcoro2020_smith} applying the expressions for the particular case of 2D space group (SG) \#14. 
The Smith normal form presented in main text can be decomposed as 

\begin{equation}
\Delta=L\cdot EBR\cdot R. 
\end{equation}
Here $EBR$ is the matrix of the elementary band representations (EBR). L and R are unimodular matrices that can be obtained from the numerical computation of $\Delta$.
Choosing a particular ordering of the irreducible representations (IR) in the EBR we have the $EBR'$ matrix

\begin{equation}
EBR' =\begin{bmatrix}
2 & 2 & 0 & 0 \\
1 & 0 & 1 & 0 \\
0 & 0 & 2 & 2 \\
1 & 0 & 1 & 0 \\
0 & 1 & 0 & 1 \\
0 & 1 & 0 & 1 \\
1 & 1 & 1 & 1 \\
1 & 1 & 1 & 1 
\end{bmatrix}.
\end{equation}

This particular choice renders all the elements in the $\Delta$ matrix positive.
Next a general band representation is defined by a vector $B$ that comprises the multiplicities of the IR of the set of bands of interest, such that 

\begin{equation}
B =\begin{bmatrix}
n_{\Gamma_{3}\Gamma_{4}}\\
n_{D_{3}} \\
n_{\Gamma_{5}\Gamma_{6}} \\
n_{D_{4}} \\
n_{D_{5}} \\
n_{D_{6}} \\
n_{Z_{2}} \\
n_{B_{2}}
\end{bmatrix}. 
\end{equation}

The topological information of the bands is encoded in the vector 
(eq.(6) in \cite{Elcoro2020_smith}) 

\begin{equation}
C=L\cdot B.
\end{equation}

In the case of SG \#14, for the particular $EBR'$ matrix presented above we obtain for the L matrix 

\begin{equation}
L =\begin{bmatrix}
0 & 0 & 0 & 0 & 1 & 0 & 0 & 0 \\
0 & 1 & 0 & 0 & 0 & 0 & 0 & 0 \\
0 & 0 & 1 & 0 & 0 & 0 & 0 & 0 \\
0 & -1 & 0 & -1 & 0 & 0 & 0 & 0 \\
1 & -2 & 1 & 0 & -2 & 0 & 0 & 0 \\
0 & 0 & 0 & 0 & -1 & 1 & 0 & 0 \\
0 & -1 & 0 & 0 & -1 & 0 & 1 & 0 \\
0 & -1 & 0 & 0 & -1 & 0 & 1 & 0
\end{bmatrix}.
\end{equation}

Thereby, the C vector is given by 

\begin{equation}
C =\begin{bmatrix}
n_{D_{5}} \\
n_{D_{3}} \\
n_{\Gamma_{5}\Gamma_{6}} \\
-n_{D_{3}}+n_{D_{4}} \\
n_{\Gamma_{3}\Gamma_{4}}-2n_{D_{3}}+n_{\Gamma_{5}\Gamma_{6}}-2n_{D_{5}}\\
-n_{D_{5}}+n_{D_{6}} \\
-n_{D_{3}}-n_{D_{5}} \\
-n_{D_{3}}-n_{D_{5}} 
\end{bmatrix}.
\end{equation}

Additionally we set $r$ as the range of the $EBR'$ matrix, that is $r=3$. The component $C_{r}$ of vector $C$ yields the definition of the topological index as stated by eq. (8) in \cite{Elcoro2020_smith}. For our case this results in the relation 

\begin{equation}
    \mathbb{Z}_2 = n_{\Gamma_{5}\Gamma_{6}} \pmod 2 ,
\end{equation}
which defines the topological invariant only in terms of the multiplicities of the $\Gamma$ point IR (An equivalent result holds for the other IR, $\Gamma_{3}\Gamma_{4}$, by choosing a different ordering in the EBR matrix).
A complementary result: using the components $C_i$ and imposing the condition $C_i = 0$ for $i>r$ allows to construct the compatibility relations for the particular band representation.  

\subsection*{B.  Spin Hall conductivity formula}

The optical spin Hall conductivity, 
 is calculated with the Kubo-Greenwood formula \cite{SHC_w90}
\begin{equation}
    \sigma_{\alpha\beta}^{\gamma}(\omega)=\frac{\hbar}{V_cN_\textbf{k}}\sum_{\textbf{k}}\sum_{n}f_{n\textbf{k}}\sum_{m\neq n}\frac{{\rm Im}\left[ \bra{n\textbf{k}}\hat{j}_{\alpha}^{\gamma}\ket{m\textbf{k}} \bra{m\textbf{k}}-e\hat{v}_{\beta}\ket{n\textbf{k}}	 \right]}
{(\epsilon_{n\textbf{k}}-\epsilon_{m\textbf{k}})^2-(\hbar\omega-i\eta)^2}.
\end{equation}
Here $\hat{j}_{\alpha}^{\gamma}=\frac{1}{2}\{ \hat{s}_\gamma,  \hat{v}_\alpha \}$ is the spin current operator with the spin operator $\hat{s}_\gamma$ defined as $\hat{s}_\gamma = \frac{\hbar}{2}\hat{\sigma}_{\gamma}$ with $\hat{\sigma}_{\gamma}$ as Pauli matrices. 
Indices $\alpha$, $\beta$ represent Cartesian coordinates and $\gamma$ appoints for the direction of spin (typically $z$-direction). $V_c$ is the cell volume and $N_k$ is the number of sampled points in reciprocal space. Finally, $f_{n\textbf{k}}$ denotes the Fermi-Dirac distribution and $\hbar\omega$ corresponds to the external perturbation frequency. The static case is calculated by taking the limit $\omega\rightarrow0$
in the above formula. 

%\newpage
\subsection*{C. Additional figures}

\begin{figure}[ht]
\centering
\includegraphics[width=1\columnwidth,clip]{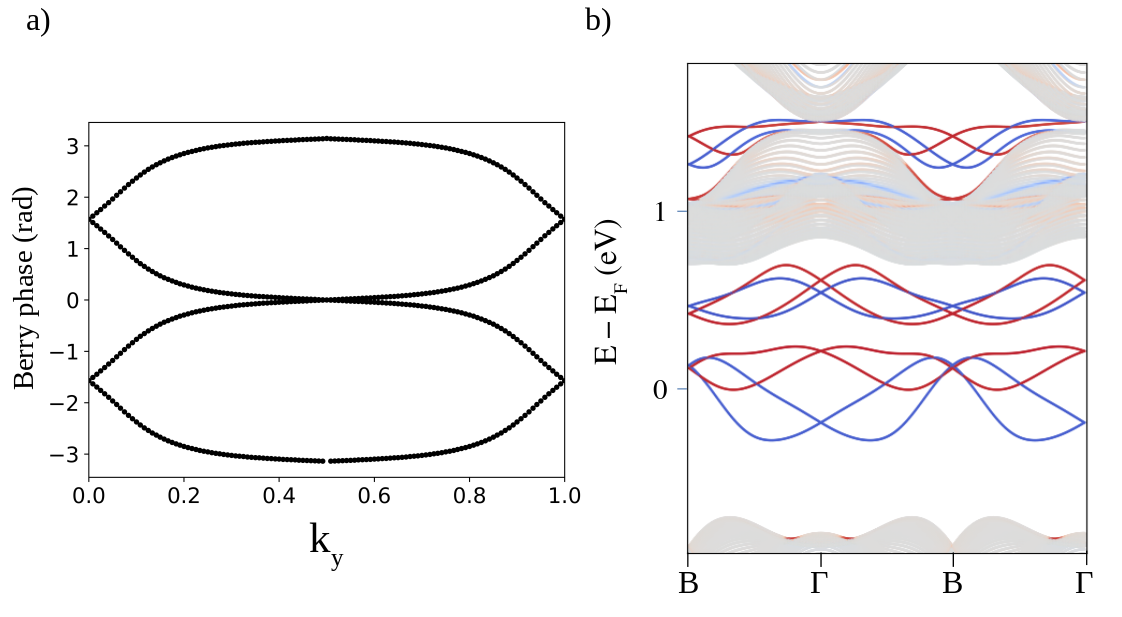}
\caption{a) Evolution of the Wannier charge center (Wilson loop) along the $k_y$ direction. 
b) Energy dispersion of the ribbons for the Wannier interpolated 12B model for a cut along the $y$ direction in real space. Red-colored and blue-colored bands represent
states that are localized at the upper and lower edges of the ribbon, respectively.}
\label{SI_1} 
\end{figure}

\begin{figure}
\centering
\includegraphics[width=1\columnwidth,clip]{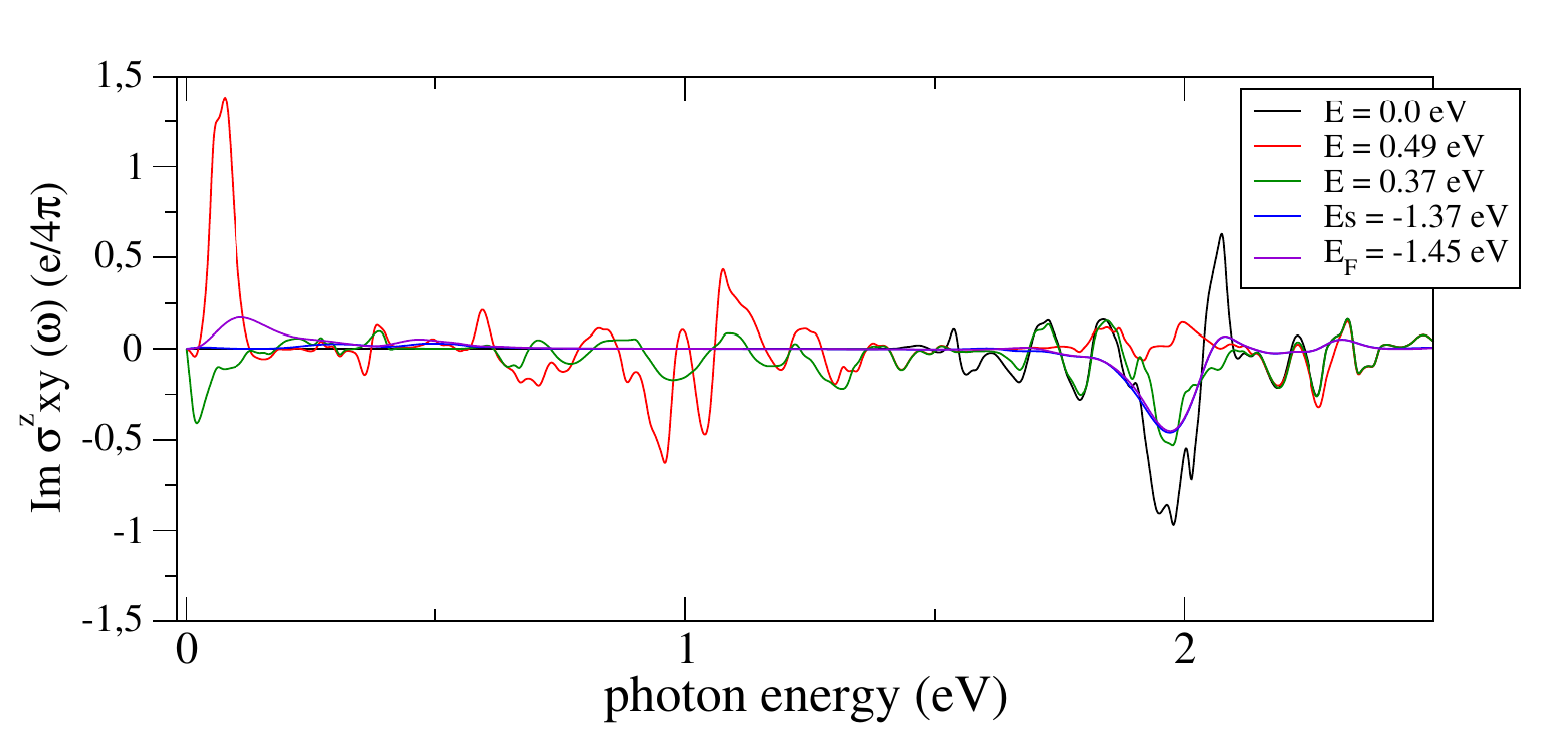}
\caption{Imaginary part of the spin Hall conductivity $\sigma^{z}_{xy}$ as a function of external photon energy for different values of the  chemical potential.}
\label{SI_2} 
\end{figure}

\begin{figure}
\centering
\includegraphics[width=1\columnwidth,clip]{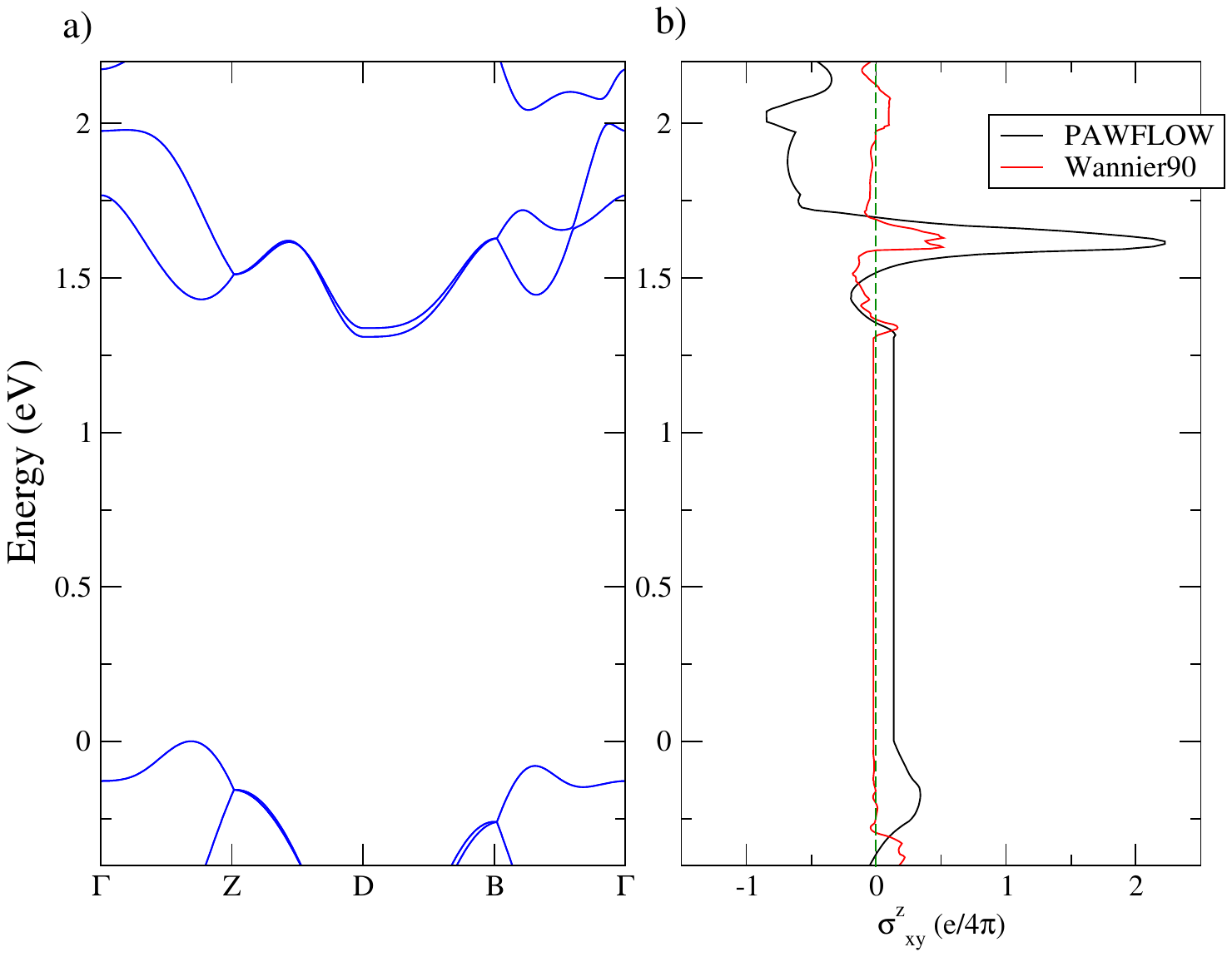}
\caption{a) Electronic bands of monolayer PdSe$_2$ from the PAWFLOW code.
b) Static spin Hall conductivity $\sigma^{z}_{xy}$ as a function of the Fermi level for two different codes.}
\label{SI_3} 
\end{figure}

\newpage
\bibliographystyle{unsrt}
\bibliography{supp}